%% file: Main.tex
\documentclass[Alon2,ChapterTOCs,singlecolor,11pt]{Alon}
\usepackage{fixltx2e,fix-cm}
\usepackage[english]{babel}
\usepackage{epigrafica}
\usepackage[LGR,OT1]{fontenc}
\usepackage{lmodern}
\usepackage{amssymb}
\usepackage{amsmath}
\usepackage{graphicx}
\usepackage{subfigure}
\usepackage{makeidx}
\usepackage{multicol}
\usepackage{lmodern}
\usepackage{amssymb}
\usepackage{amsmath}
\usepackage{graphicx}
\usepackage{subfigure}
\usepackage{makeidx}
\usepackage[algoruled]{algorithm2e}
\usepackage{multicol}
\usepackage{caption}
\usepackage{changepage}% for text width to include margin
\usepackage{version}
\title{Alon Template}
\begin{document}
\tableofcontents
\include{Chapters/Chap1/ch}

\end{document}

%% file: Chapters/Chap1/ch.tex
\chapterauthor{Uttam Ghosh}
{Department of EECS, Vanderbilt University, Nashville, TN, USA-37235}
\chapterauthor{Pushpita Chatterjee}
{Department of Computational, Modeling and Simulation Engineering, Old Dominion University, Suffolk, VA 23435}
\chapterauthor{Sachin Shetty}{Department of Computational, Modeling and Simulation Engineering, Old Dominion University, Suffolk, VA 23435}
\chapterauthor{Raja Datta}{Dept. of Electronics and Electrical Communication Engineering, Indian Institute of Technology Kharagpur, Kharagpur, West Bengal, India 721302}

\chapter[SDN-IoT based Smart City Framework]{An SDN-IoT-based Framework for Future Smart Cities: Addressing Perspective}

\chapterinitial{I}n \emph{this Chapter}, a software-defined network (SDN)-based framework for future smart cities has been proposed and discussed. It also comprises a distributed addressing scheme to facilitate the allocation of addresses to devices in the smart city dynamically. The framework is dynamic and modules can be added and omitted by a centralized controlling unit without disturbing the other components of the framework and other modules may be updated accordingly. In the proposed addressing scheme, a new Internet of Things (IoT) device will receive an IP address from one of their existing neighboring peer devices. This allows devices in the city to act as a proxy and generate a set of unique IP addresses from their own IP addresses, which can then be assigned to new (joining) devices; hence, reducing addressing overhead and latency, as well as avoiding the need to send broadcast messages during the address allocation process. Thus, it achieves considerable bandwidth and energy savings for the IoT devices. 

\section{Introduction}\label{Intro}
It has been estimated that approximately 65\% of the world's population will eventually live in cities by the year 2040 \cite{b1}. There has been a trend of making cities smarter, for example by leveraging existing and emerging technologies such as Internet of Things (IoT). The latter can be broadly defined to be a (heterogeneous) network of a broad range of physical Internet connected devices, such as smart vehicles, smart home appliances, and other devices with embedded software or hardware (e.g., sensors), that can be used to connect, sense / collect, and disseminate / exchange large volume of data. This also allows us to offer advanced services that can be used to improve the quality of service delivery and life. 

The increasing trend of smart cities is partly due to the lower of technological and cost barriers in deploying communication networks (e.g. wireless and 5G) in a broad range of settings, such as residential and commercial buildings, utility networks, transportation networks, and those in the critical infrastructure sectors \cite{b2,b3}. In such settings, it is clear that data plays a key role, for example in informing decision and strategy making and formulating. Such data can be collected by the broad range of IoT devices and networks, and can be compiled and analyzed to achieve improved service delivery in healthcare, manufacturing, utility, supply chain and many other services. However, there exist a number of challenges in dealing with such data, due to the volume, variety, velocity, and veracity (also commonly referred to as the four V's of big data). For example, the management and performance optimization of IoT-based smart cities and programmability of things can be extremely complex, and also the inter-connectivity can introduce security implications. Therefore, how to ensure that the underpinning communication infrastructure in the smart city is scalable, reliable, secure and efficient can be challenging, both operationally and research-wise. 

Emerging software-defined networking (SDN) decouples the control plane and data plane and subsequently it enables the control plane to become directly programmable and the underlying infrastructure to be abstracted for the applications and the network services. SDN controller, also called network operating system (NOS), is logically centralized and responsible for controlling, managing, and dynamically configuring the devices at the data plane of the network. It is effective in taking decisions for the routing, quality-of-service (QoS) and load balancing dynamically. It is easy to add new network functionalities through application programs due to the programmability feature of SDN controller. Moreover, SDN enhances the network performance by providing security and the network virtualization features. SDN controller is capable to monitor all the nodes and their traffic, and eliminate the attacker node from the network on-fly by writing effective flow rules on the switches at data plane \cite{chapter-crc}.  

\noindent \textbf{Motivation:} 
Each device in the infrastructure should have a unique address by which it can be identified. This unique address enables unicast communication and routing between devices in the infrastructure. However, as more IoT devices are introduced in the smart city, the demand for these unique addresses increase rapidly. Manual configuration of IoT devices in most of the cases inapplicable and error prone due to large size of the network. Further, centralized Dynamic Host Configuration Protocol (DHCP) \cite{b9} is not a suitable solution as the sever has to maintain configuration information of all the nodes in the network. 

\begin{figure}[!ht]
\centering
\includegraphics[height=0.48\textwidth]{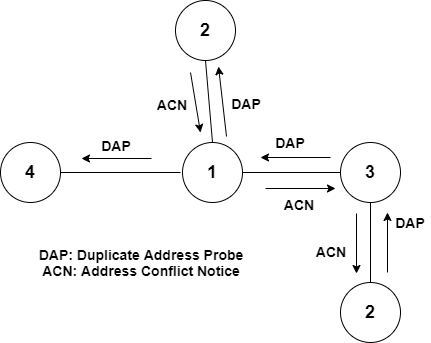}
\caption{Duplicate Address Detection (DAD) mechanism} 
\label{dad}
\end{figure}

Duplicate Address Detection (DAD) mechanism \cite{b10} can be used to resolve address conflict in the smart city. In DAD, a joining node chooses a tentative IP address randomly and verifies the whether this address is available for use or not. In order to verify the uniqueness of the address, the joining node floods a Duplicate Address Probe (DAP) message throughout the smart city and starts a timer to receive Address Conflict Notice (ACN) message from the network. If no ACN message is received, then the joining node concludes that the tentative address is free to use and configures itself with the address permanently. It has to run the DAD process again in case the joining node receives a ACN message from the network. The addressing overhead for DAD mechanism is very high as it needs to flood a message throughout the network. Further, the broadcast storm problem \cite{b11} can be seen in DAD. Figure \ref{dad} shows the DAD mechanism where a new node tries to join the network.

\textbf{Contribution:} It can be seen from the above discussion that there is a need to design a distributed addressing scheme to efficiently handle the ever increasing requirement in SDN-IoT based smart city networks. Further, the addressing scheme should assign unique IP addresses to the devices of the network for the correct routing and unicast communications. Furthermore, the scheme needs to be scalable and should not degrade its performance with respect to addressing overhead when the network size is very large like a smart city. This Chapter has two significant contributions: 

\begin{itemize}
\item firstly, an SDN-based IoT framework for a smart city architecture,
\item and secondly, a distributed addressing scheme to efficiently assign a unique IPv6 address to each device in the proposed smart city framework.
\end{itemize}
With this Chapter, readers can have a more thorough understanding architectures of SDN, IoT, and SDN-IoT-based smart cities. It further proposes an IPv6 addressing mechanism to allocate unique address to each IoT devices in a SDN-IoT-based smart city.

\noindent \textbf{Chapter Organization:} The rest of the Chapter is organized as follows: Section \ref{background} presents a background of software-defined networking (SDN), Internet of Things (IoT) and IPv6 addressing. Section\ref{rel} discusses state-of-the-art literature on SDN-IoT based networks and also address allocation techniques in various wireless networks. The proposed framework for SDN-IoT-based smart city with an addressing scheme is presented in Section\ref{proto}. Finally, Section \ref{conclu} concludes the Chapter.

\section{Background}\label{background}
In this section, we give an overview of basic preliminary concepts of Software-defined Networking (SDN), Internet of Things (IoT)  and IPv6 addressing. 
\begin{figure}[ht]
\centering 
\includegraphics[height=0.5\textwidth]{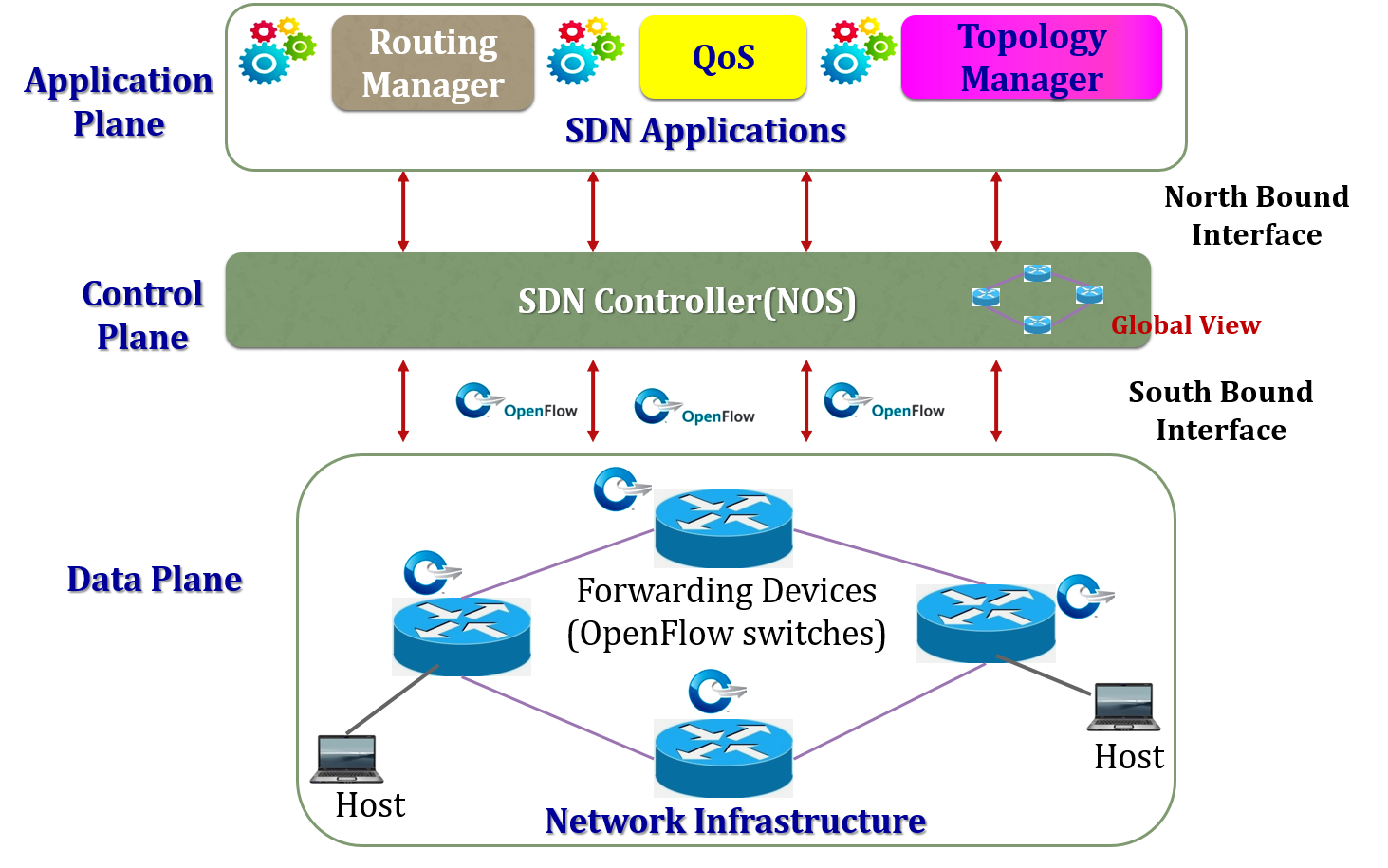}
\caption{A layering architecture of SDN} 
\label{SDN-arch}
\end{figure}

\subsection{An Overview of SDN} 
This Section presents an overview of SDN architecture and its working principles. It also presents the need of SDN and how SDN is different as compared to the traditional networking. Figure \ref{SDN-arch} presents the major elements, planes (layers) and interfaces between layers of SDN architecture. It has three planes: data plane, control plane and application plane.

\begin{figure}[!ht]
\centering 
\includegraphics[height=0.5\textwidth]{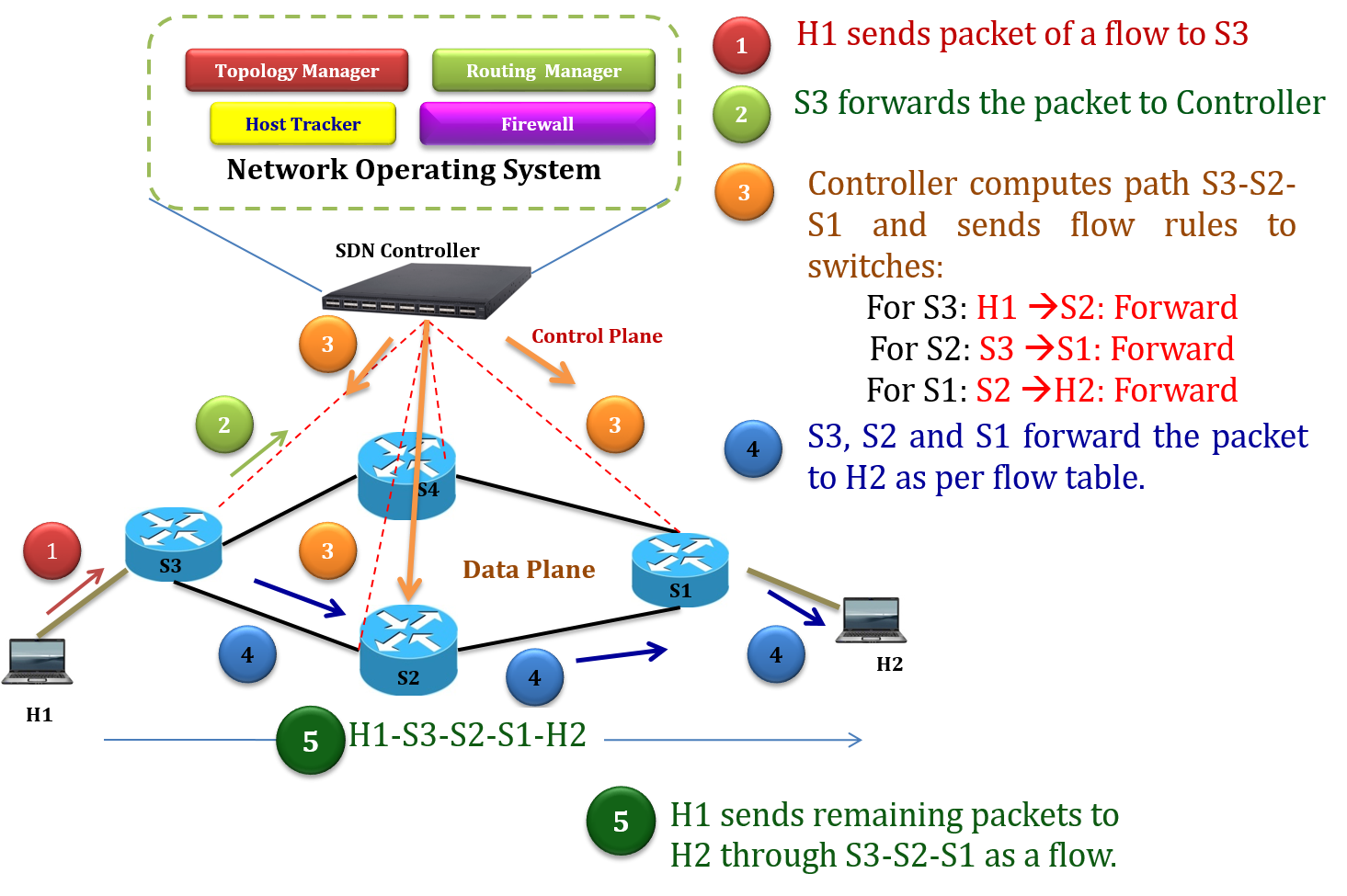}
\caption{Working Principles of SDN} 
\label{working-principle}
\end{figure}

\textbf{Data Plane:} The first plane in SDN architecture is the data plane (also known as infrastructure plane) that consists of hosts and traffic forwarding devices. These traffic forwarding devices are known as OpenFlow (OF) switches. These switches are called dump switches and able to forward the data from source host to destination host only after receiving the instructions (flow rules) from the SDN control layer.

\textbf{Control Plane:} The second plane in SDN architecture is the control plane that may comprise an SDN controller or a set of SDN controllers. SDN controller (also called network operating system (NOS)) is a logical entity (software programs) which is programmable. It is logically centralized. Hence it can track the network topology (global view of the network) and the statistics of the network traffic periodically. Further, SDN controller is
responsible for controlling, managing, and dynamically configuring the devices at the data plane of the network. It efficiently provides routing, quality-of-services (QoS), security and also balances the load in the network.  

\textbf{Application Plane:} The third and final plane is the application plane in SDN architecture. This plane runs application programs and uses application programming interface (API) to control the network resources with the SDN controller. These application programs periodically collect information from SDN controller and provide services (e.g., routing, quality of services (QoS) and load balancing). This plane also provides a programming interface to the network administrator for developing applications according to the requirements of the network. For instance, an application can be built to monitor all the devices and their traffics periodically for detecting the misbehaving devices in the network.

The northbound application programming interface (API) defines the connection between application plane and control plane whereas the southbound API defines the connection between control plane and date plane. OpenFlow (OF) protocol has been widely used as the southbound API. The SDN controller uses OpenFlow protocol to send the flow rules to the OpenFlow switches in data plane. OpenFlow protocol uses secure socket layer (SSL) and TCP for providing secure communication and reliable delivery of data between the controller and OF switches respectively.

\begin{figure}[!ht]
\centering 
\includegraphics[height=0.5\textwidth]{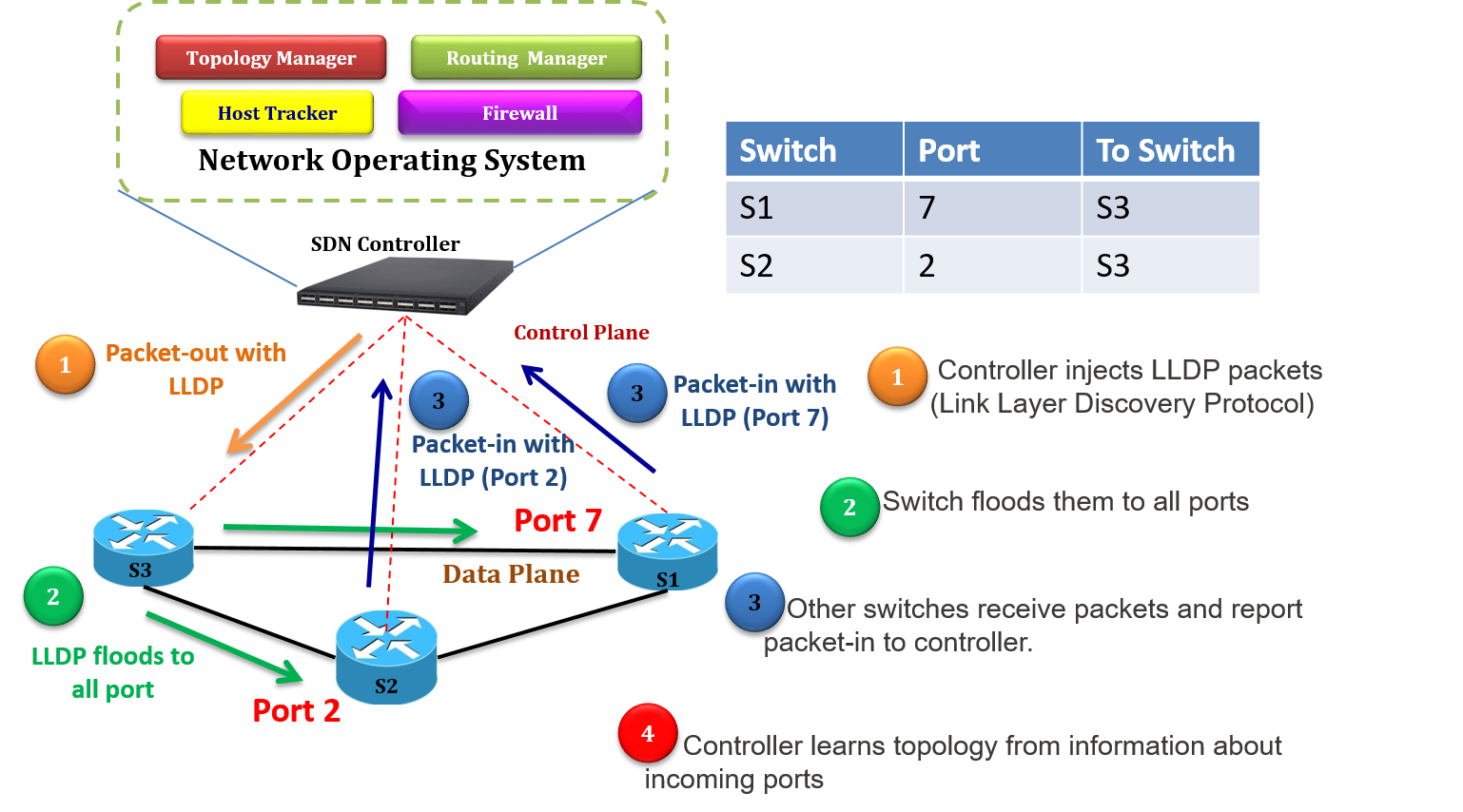}
\caption{Topology detection using LLDP} 
\label{LLDP}
\end{figure}

The working principle of SDN is presented in Figure \ref{working-principle}. A device H1 (source) sends the packets of a flow to another device H2 (destination) through OF switches S3-S2-S1 in an SDN-based network \cite{chapter-crc}. Here, the SDN controller detects topology of the network using link layer discovery protocol (LLDP) as shown in Figure \ref{LLDP}. Thus, it knows the global topology of the network and responsible for the routing between the devices. 

\begin{figure}[!ht]
\centering 
\includegraphics[height=0.5\textwidth]{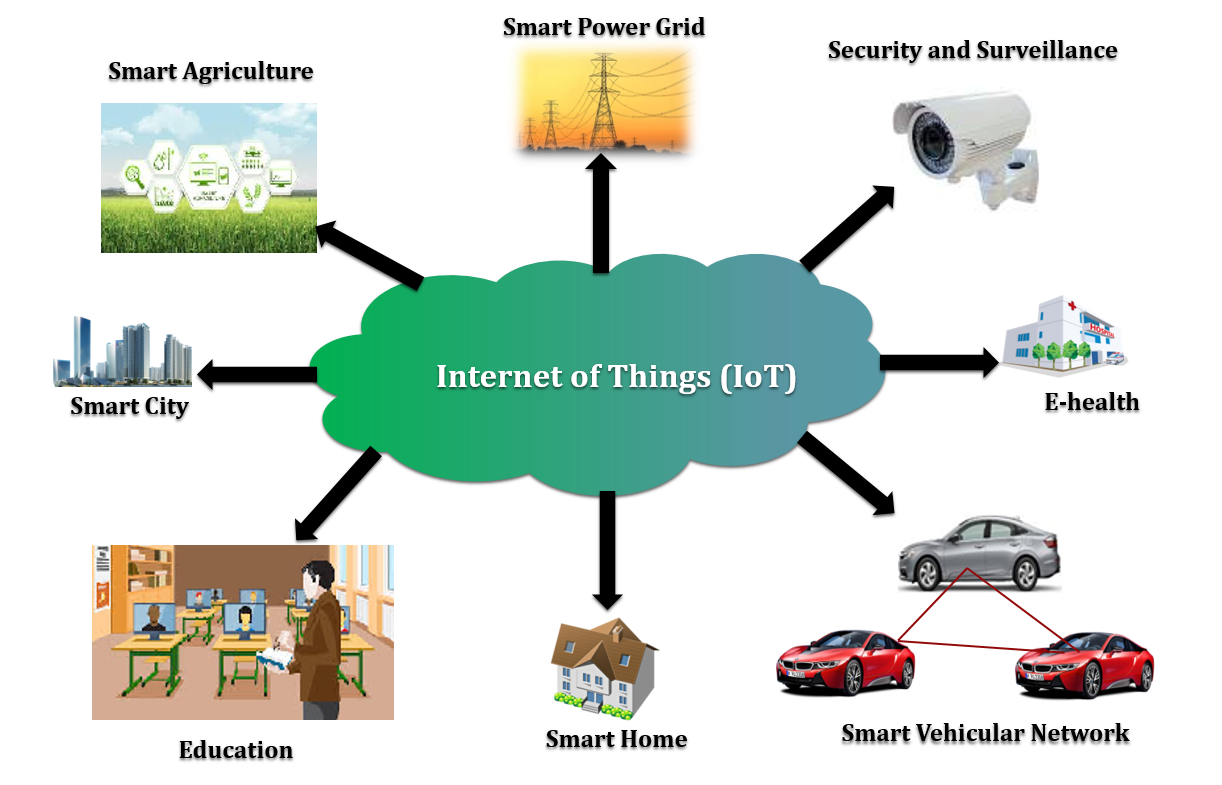}
\caption{Internet of things (IoT)} 
\label{IoT-basic}
\end{figure}
\subsection{An Overview of IoT and Smart Cities}
\noindent \textbf{The Internet of Things (IoT)}:  An IoT is a heterogeneous network of physical objects (things) that are embedded with electronics, sensors, software, actuators, RFID tags, and other technologies for connecting and communicating a large amount of data with other devices and networks over the Internet to offer a new class of services at anytime, anywhere and for anyone. It can form a large network by combining wired networks and different types of wireless networks such as wireless sensor networks (WSNs), ZigBee, WiFi, mobile ad hoc networks (MANETs), and RFID. IoT can be applied to make the physical infrastructures more smart, secure and reliable, and fully automated systems. These physical infrastructures include buildings (homes, schools, offices, factories, etc.), utility networks (gas, electricity, water, etc.), healthcare systems, transportation vehicles (cars, rails, planes, etc.), transportation networks (roads, railways, airports, harbors, etc.), and information technology networks, etc. IoT collects, stores, and exchanges a large volume of heterogeneous data from various types of networks and provides critical services in smart homes and buildings, healthcare systems, transportation networks, utility networks, industrial control and monitoring systems, and so on \cite{chapter-crc,b12c,b12a,b12b}.

\begin{figure}[!ht]
\centering 
\includegraphics[height=0.5\textwidth]{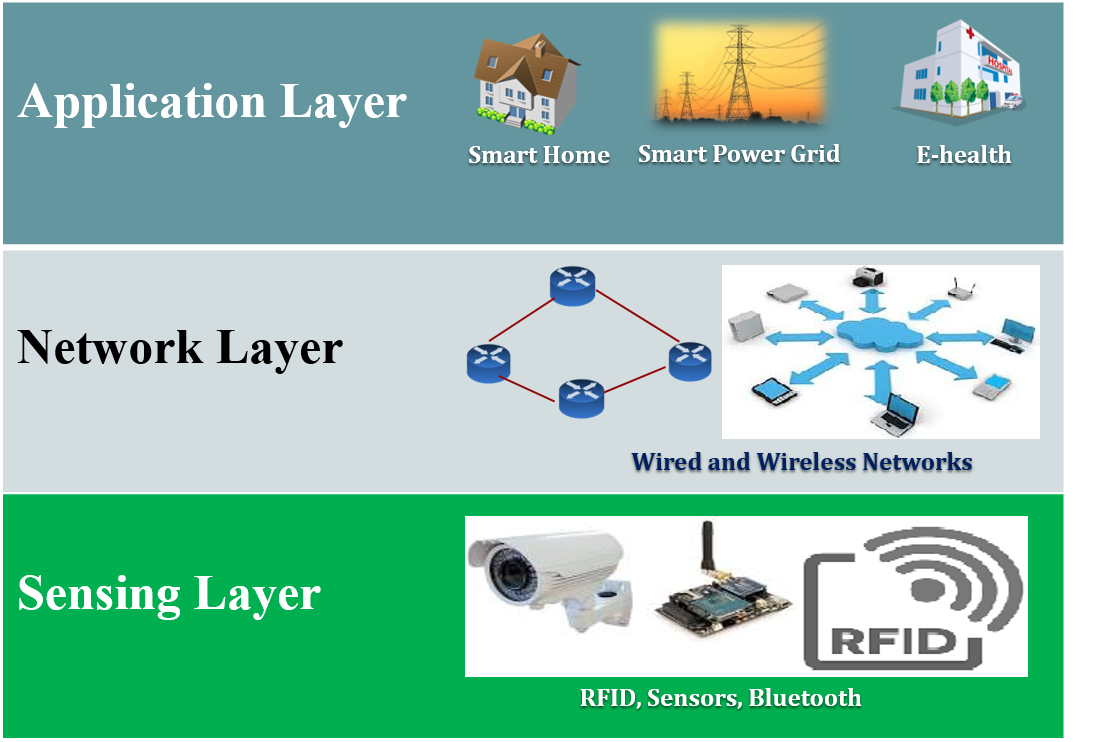}
\caption{The three-layered architecture of IoT} 
\label{IoT-layers}
\end{figure}

Figure \ref{IoT-layers} shows the layering architecture of IoT. It comprises of three main layers: sensing layer, network layer and application layer. The sensing layer, also known as a perception layer, consists physical objects and sensing devices. This layer is responsible for sensing and collecting the data from the physical objects. Network layer bridges between sensing layer and application layer. It carries the data collected from the physical objects through sensors. The network can be wireless or wired network for the transmission. Thus, network layer is responsible for connecting the smart things, network devices and networks to each other and also for transmitting the data from physical objects to the gateway of the network. Application layer is responsible for providing the services to the users based on their demands and applications. The applications of IoT can be smart homes and buildings, smart grids, smart health, smart cities, etc.

\begin{figure}[!ht]
\centering 
\includegraphics[height=0.5\textwidth]{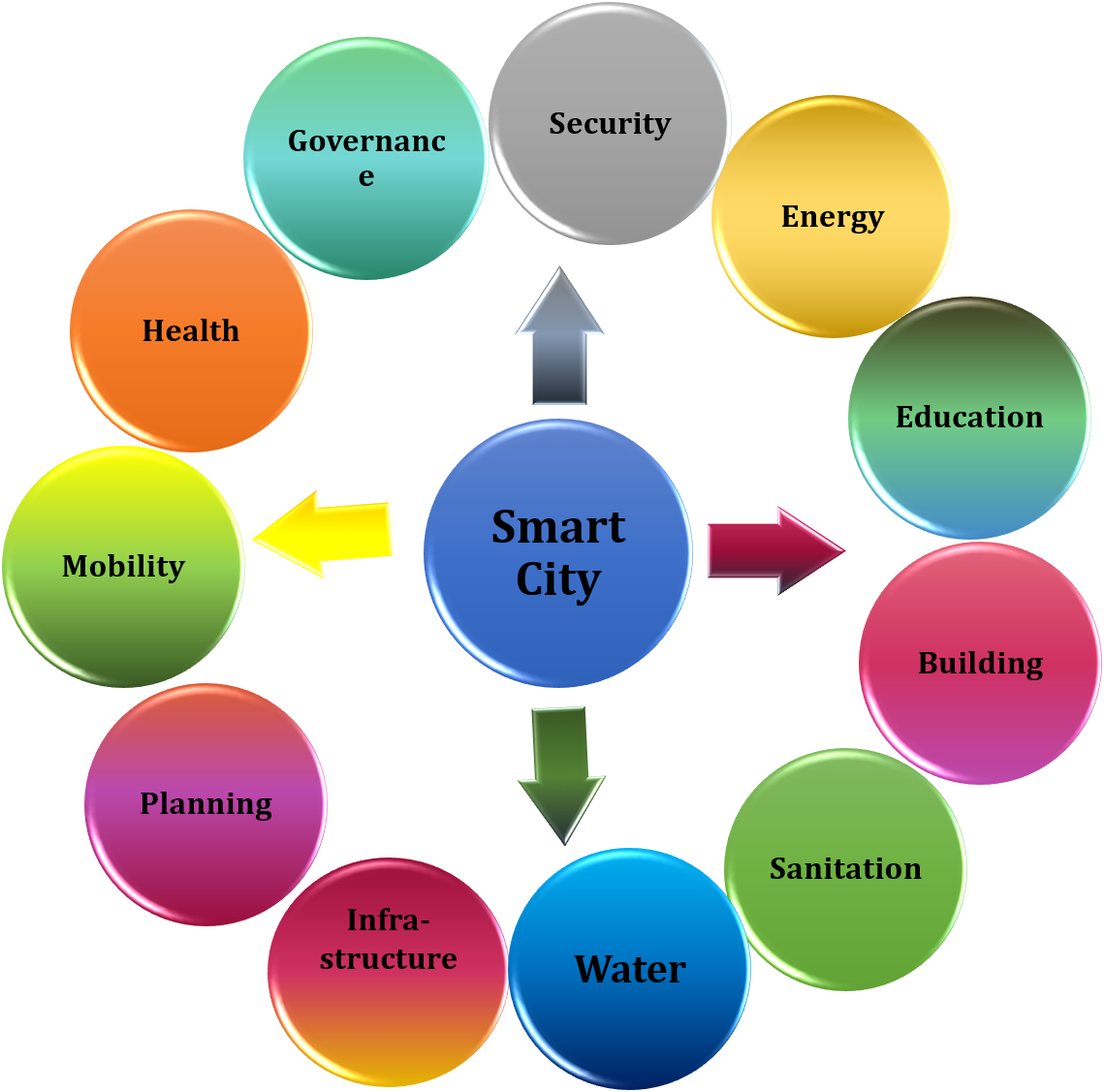}
\caption{Overview of Smart city components} 
\label{overview}
\end{figure}

\par \noindent \textbf{Smart City}: A smart city is an urban area that uses different types of IoT devices to collect, process and analyze the data for monitoring and managing traffic and transportation systems, utilities, power grids, waste management, water supply networks, schools, libraries, hospitals, security and surveillance systems, and other community services. It helps city officials to interact directly with both community and city infrastructure and also to monitor and manage the city resources efficiently and smartly. The main components of a smart city is depicted in Figure \ref{overview}. 

\subsection{An Overview of IPv6 Addressing} 
\begin{figure}[ht]
\centering 
\includegraphics[height=0.5\textwidth]{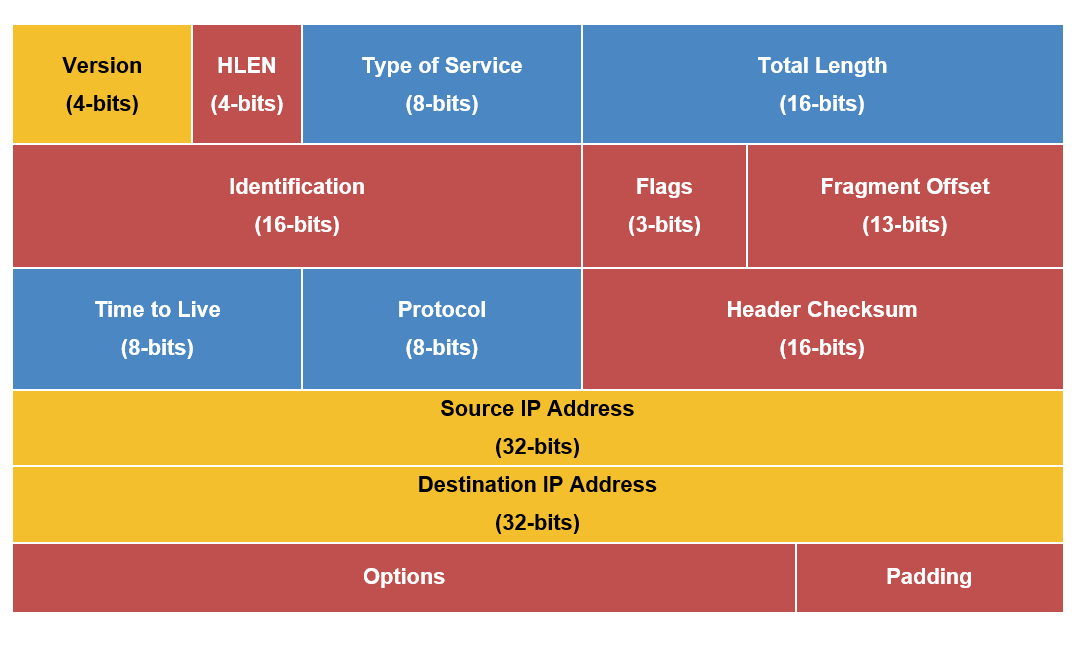}
\caption{IP version 4 (IPv4) Header Format} 
\label{IPv4}
\end{figure}

\begin{figure}[ht]
\centering 
\includegraphics[height=0.5\textwidth]{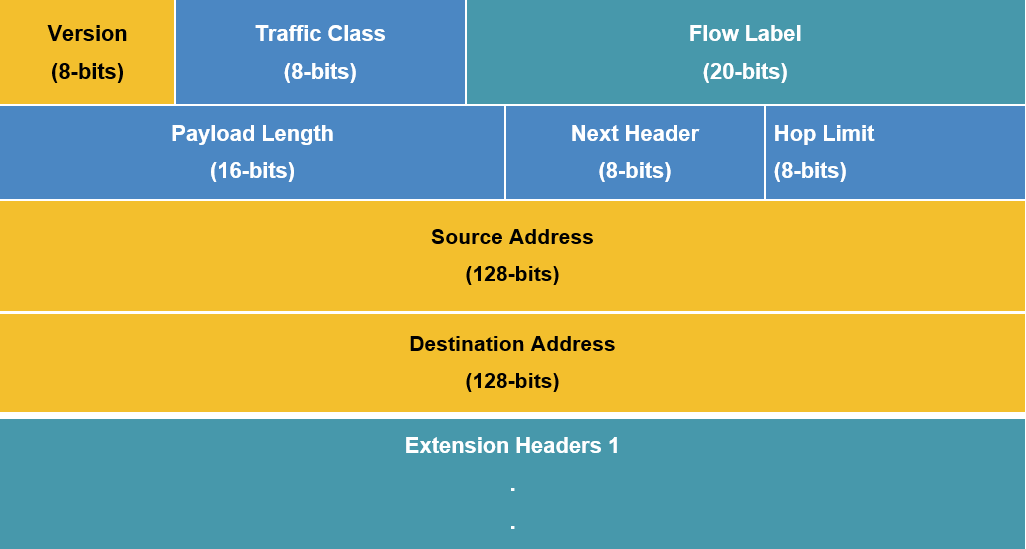}
\caption{IP version 6 (IPv6) Header Format} 
\label{IPv6}
\end{figure}

Internet protocol version 4 (IPv4) is the most widely deployed IP used to connect devices to the Internet. IPv4 addresses are 32-bit long and can be used to assign a total of $2^{32}$ devices (over 4 billion devices) uniquely.  However, with the growth of the Internet and IoT it can be expected that the number of IPv4 addresses may eventually run out as each device that connects to the Internet and IoT requires an IP address. A new IP addressing system Internet Protocol version 6 (IPv6) is being deployed to fulfill the need for more IP addresses. An IPv6 addresses are 128-bit long and can be used to assign a total of $3.4 * 10^{38}$ devices uniquely. Further, it supports auto-configuration and provides better quality of services (QoS), mobility and security as compared to IPv4. Figure \ref{IPv4} and Figure \ref{IPv6} present the headers of IP version 4 and IP version 6 respectively.

\noindent \textbf{IPv6 Address Representation:} An IPv6 address is represented as eight groups of four hexadecimal digits where each group represents 16-bits. These groups are separated by colons (:). An example of an IPv6 address is:

\begin{center} 2031:0000:130f:0000:0000:09c0:876a:130b \end{center}

Leading zeroes in a group are optional and can be omitted. One or more consecutive groups containing zeros can be replaced by double colons (::), but only once per address. Therefore, the example address can be written as:

\begin{center} 2031:0:130f::9c0:876a:130b \end{center}

\noindent \textbf{IPv6 Header Format:} The header format of IPv6 is shown in Figure \ref{IPv6}. Here, the fields of IPv6 header have been discussed briefly: 

\noindent \textit{Version:} This field indicates the version of Internet Protocol which contains bit sequence 0110.

\noindent \textit{Traffic class:} This field presents the class or priority of IPv6 traffic as it is similar to service field in IPv4 header. The router discards the least priority packets if congestion occurs in the network.

\noindent \textit{Flow label:} Source node uses flow label field to label the packets belonging to the same flow in order to request special handling (for example, quality of service or real time service) by intermediate IPv6 routers. It also specifies the lifetime of the flow.

\noindent \textit{Payload length:} This field indicates the total size of the payload including extension headers (if any) and upper layer data.

\noindent \textit{Next header:} This filed is used to indicate the type of extension header (if any) immediately following the IPv6 header. It also specifies the upper-layer protocols (UDP, TCP) in some cases.

\noindent \textit{Hop limit:} This field is same as time-to-live (TTL) field in IPv4 header. It specifies the maximum number of routers an IPv6 packet can travel. The value of the hop limit gets decremented by one by each router that forwards the packet. The router discards the packet if the value of the hop limit reaches to 0. This filed prevents the packet from circulating indefinitely in the network.

\noindent \textit{Source address:} This filed specifies the IPv6 address of the original source of the packet.

\noindent \textit{Destination address:} This filed indicates the IPv6 address of the final destination. In order to correctly route the packet, the intermediate routers use destination address of the packet.

\noindent \textit{Extension header:} This field have been introduced to allow the incorporation and usage of several options whenever is needed. The size of the IPv6 main header is 40-bytes long. Next Header field of IPv6 main header points to the first Extension Header and the first extension header points to the second extension header and so on.

\section{Related Works}\label{rel}
A number of different approaches have been explored in the literature, including the use of software-defined networking (SDN). For example, there have been attempts to integrate SDN and IoT technologies into the heterogeneous communication infrastructure in smart cities \cite{b4,b5,b6,b7}, by say  utilizing SDN to manage and determine the correctness of network operations at run-time. This is because we can leverage the globalized view and the programmability features available in the SDN controller to control, configure, monitor and detect faults and mitigate abnormal operation(s) in the underpinning infrastructure; hence, allowing us to achieve efficiency and reliability.
\par Mavani et al., has done several works on secure addressing and privacy preserving methodologies for IoT and Mobile environment paradigm \cite{b13,b14,b15}. In IoT, billions of devices can be addressed using IPv6 addressing scheme. Attackers can spoof addresses from unsecure wireless communication channels and advertise them as a legitimate device.  Malicious users can track activity of these devices by spoofing IPv6 addresses. To mitigate this type of attacks by hiding the IPv6 address from attacker. They have proposed a secure privacy preserving method\cite{b13}, which changes the IPv6 address of each device periodically and pseudorandomly in order to hide its identity. They analyzed the method using Cooja simulator to show that the method does not inflict much overhead for random changing of address and reconfiguration. In \cite{b14, b15}, they investigated the use of secure addressing and privacy mechanisms for f IPv6 over Low -Power Wireless Personal Area Networks (6LoWPAN) and designed a method to provide resilience against address spoofing and better reconfiguration time from attack disruption. They showed the efficacy of their proposal by time complexity analysis and simulation with benchmark data, but overall this does not pose much overhead to provide resilience against address spoofing.  
\par Brilli et al., proposed a secure privacy aware two layer addressing scheme for 6LoWPAN wireless network in order to improve security and privacy along with reducing the chance of spoofing by hiding the traceability of the user \cite{b16}. With a minimal overhead and using standard 6LoWPAN messages security and privacy have been ensured in an energy constrained environment. 
Wang et al., proposed a long-thin and tree-based topology in addressing-based routing optimization in vehicular scenarios (AROV) \cite{b17}  to provide unique address to sensor nodes in 6LoWPAN wireless sensor networks Using a concept of super node for multi-hop sensor nodes serves as address initiator for its all neighbor nodes. They have shown it mitigates address failure and also gives performance in routing by reducing latency. The authors also proposed location aware addressing for 6LoWPAN wireless sensor networks \cite{b18}. In this addressing scheme  without using duplicate address detection, a node can obtain a globally unique address. The address initialization is done zone wise where zones are independent of the one in another. therefore this parallel and address initialization took less time. Wang et al, further proposed stateful address configuration mechanism for wireless body area networks \cite{b19}. The uniqueness of the address is maintained without duplicate address detection. Automatic reclamation of unused or released address have been done without any extra overhead. Using simulation they have shown the efficacy of performance by reducing the address configuration delay and cost.
For heterogeneous wireless network a dynamic Internet Protocol (IP) address assignment architecture \cite{b20} has been proposed by Khair et al. The addressing mechanism introduced security and service reliability with a reduced Opex.  However, this scheme does not perform well in heterogeneous heavy traffic scenarios as it incurs significant overhead.
Li et al.,  presented address configuration algorithm for network merging in Ad hoc network scenario \cite{b21}. By restricting the new address generation only duplicate addresses during merging their scheme significantly improve the network performance. 

In \cite{b22}, an IP-based vehicular content-centric networking framework has been proposed by Wang et al., by employing the unicast address-centric paradigm to achieve content acquisition. They avoid the broadcast centric communication.  Using the unicast communication, they have shown it substantially reduces the content acquisition cost and gives better performance in success rate content acquisition. 

In \cite{b24}, El-Shekeil et al., investigated several conflict scenarios of using Private IP for enterprise network.  They formulated the problem to minimize the routing table sizes  as NP-Hard. They devised effective heuristics formulation in order to solve the problem. To prove the efficacy of the same they provided empirical result which showed significant reduce in the number of subnet entries and the routing table sizes.

A Mobile Ad-hoc Network (MANET) is a collection of mobile nodes with a dynamic self-configured network. It has no fixed and pre-established infrastructure without any centralized administrations or base stations. MANET can be integrated with IoT to implement smart cities. Therefore, IP addressing is very important and challenging issue for a MANET as it is an infrastructure-less and highly dynamic network.  In light of this, Ghosh et al., proposed IPv6-based and IPv4-based secure distributed dynamic address allocation protocols \cite{b24b, b24c, b25,b25a, b25b, b26}. In these protocols, the new node gets an IP address from its neighbors acting as proxies. The new node becomes proxy once it receives an IP from the network. Further, these protocols can handle the network events such as network partitioning and merging without using complex duplicate address detection mechanisms. 

Akhtar et al., proposed a congestion avoidance algorithm \cite{b26a} for IoT-MANET which used bandwidth as the main component to find the optimal route.  By getting feedback about  the residual bandwidth of network path each channel aware routing scheme (BARS) that can avoid congestion by monitoring residual bandwidth capacity in network paths they significantly improve network parameters like of latency, end-to-end delay  and packet delivery ratio for both static and  dynamic network  topologies. A secure SDN based framework has been proposed for content centric application has been devised by Ghosh et al. In \cite{b27}, secure multi-path routing protocol has been designed which significantly improves the network performance.This work is pretty much feasible to incorporate for futuristic smart cities. Ghosh et al., proposed a SDN based secure framework for smart energy delivery system \cite{b28}or smart cities, which addressed a number of fault injections and controller failure scenarios as well. In \cite{b29}, Alnumay et al., designed and developed a trust-based system for securing IoT applications using a predictive model of ARMA/GARCH (1,1), whcih significantly improve network functionalities in smart city scenarios. 

\section{The Proposed SDN-IoT-based Smart City Framework}\label{proto}
\begin{figure}[!ht]
\centering 
\includegraphics[height=0.5\textwidth]{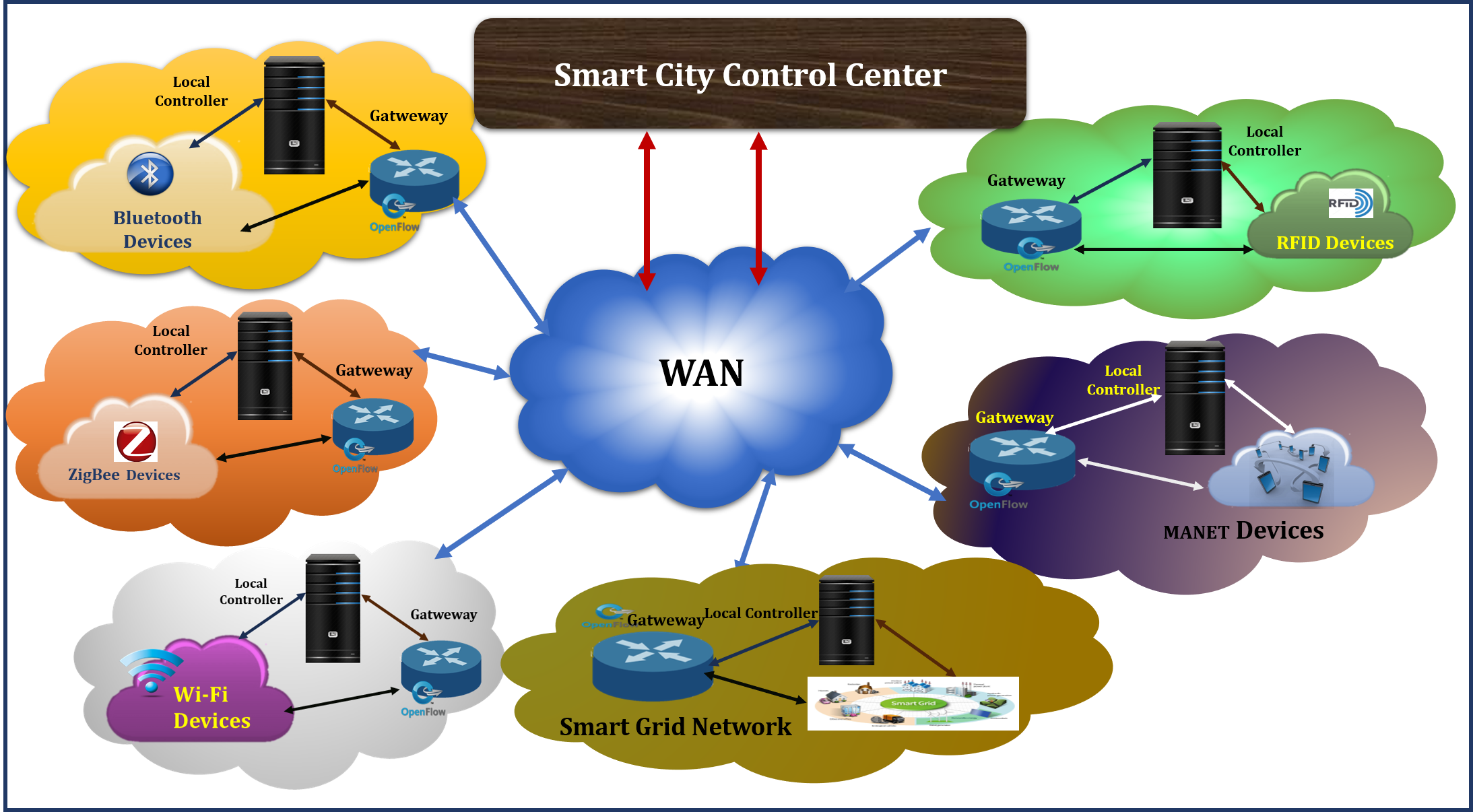}
\caption{An SDN-IoT-based smart city framework} 
\label{Smartcity}
\end{figure}

Here, we propose our SDN-IoT based smart city framework, which is configured, controlled, and managed by a global control center as shown in Figure \ref{Smartcity}. The proposed framework supports heterogeneous networks and contains different types of networks including ZigBee, mobile ad-hoc networks (MANETs), sensor networks and Bluetooth. 

\begin{figure}[!ht]
\centering 
\includegraphics[height=0.6\textwidth]{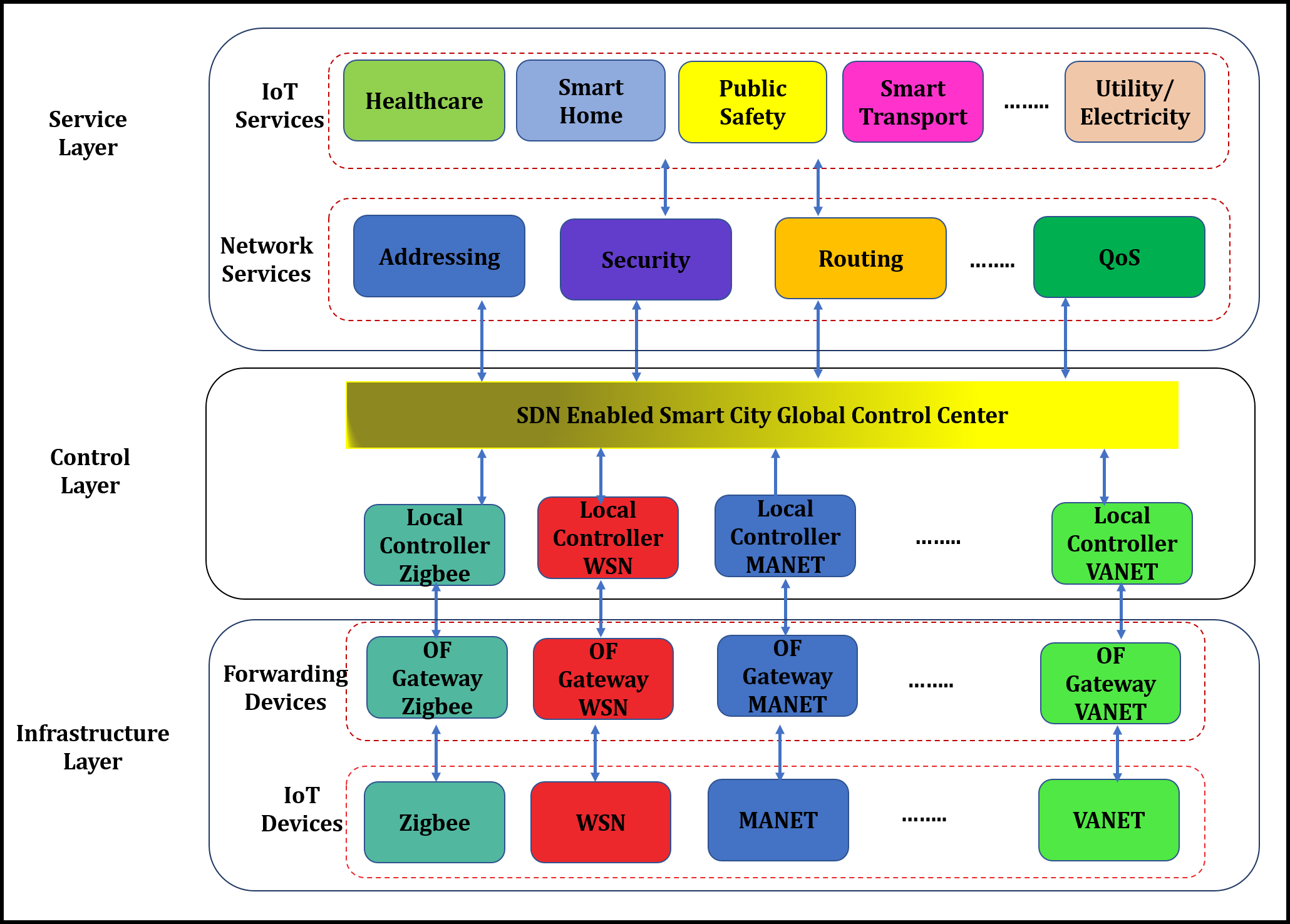}
\caption{An SDN-IoT-based layered smart city framework} 
\label{arch}
\end{figure}
We also present a SDN-IoT based layered smart city framework in Figure \ref{arch}. Our proposed architecture has three layers, described as follows. The first layer is the infrastructure layer, which consists the IoT devices sublayer and the forwarding devices sublayer. The IoT devices sublayer contains different types of wireless devices (e.g. ZigBee, sensors, and Bluetooth) to create different types of IoT application domains. These wireless devices collect large volume of data from the networks and send them to the global smart city control center for further processing. The IoT device sublayer also contains actuators to receive control commands from the global control center and execute them. The forwarding devices sublayer consists of Openflow (OF) gateways, which facilitate the forwarding of control and data packets to the global control center. The control layer contains a global SDN controller and a number of local SDN controllers. The global SDN controller is mainly responsible for controlling and monitoring  communications between global control center to IoT application domains and an application domain to other application domains, and the local SDN controller controls and monitors the communication between devices inside an application domain. The application layer provides IoT services (e.g. smart homes, smart grids, and smart transportation) using SDN controllers. It further provides network services such as routing, security and quality of service (QoS) in the city. 

\subsection{The Proposed Addressing Scheme} 
Here, we discuss our proposed IPv6 addressing scheme that is designed to provide unique addresses to IoT devices in the infrastructure. Using the proposed addressing scheme, unique IP addresses can be generated from the IP address of an existing device in the city (network), which can then be provided to new / joining IoT devices. In other words, without the need to broadcast any message over the entire city, any new / joining IoT device can acquire an IP address from its peers / neighboring devices. This concept is adopted from \cite{b8}. 

Here, we discuss the algorithm given in Function
ip-generation that generates unique IPv6 addresses
for new IoT devices joining the network. As discussed, an IPv6 address comprises eight (8) groups of four (4) hexadecimal (\textsl{HEX})
digits, which are separated by colons (for
example, 2031:0000:130f:0000:0000:09c0:876a:130b). The IPv6 address logically divided into two parts: a 64-bit network prefix and a 64-bit interface identifier. For ease of presentation, we express the address in 16-byte dotted decimal (\textsl{DEC}) format: ($b_{15}.b_{14}.b_{13}.b_{12}.b_{11}.b_{10}.b_9.b_8.b_7.b_6.b_5.b_4.b_3.b_2.b_1.b_0$)$_{DEC}$ wherein  $b_{15}.b_{14}.b_{13}.b_{12}.b_{11}.b_{10}.b_9.b_8$ and $b_7.b_6.b_5.b_4.b_3.b_2.b_1.b_0$ are the network prefix (which is fixed for a network domain) and the device identifier respectively.

\begin{function}%[!ht]
%\SetAlgoLined
%\footnotesize {
\scriptsize{
%\SetLine
getmyip $\leftarrow (b_{15}.b_{14}.b_{13}.b_{12}.b_{11}.b_{10}.b_9.b_8.b_7.b_6.b_5.b_4.b_3.b_2.b_1.b_0)_{DEC}$\;
Set $static \ count \leftarrow 0$, $count1 \leftarrow 1$,$j \leftarrow 0$, $i \leftarrow 0$\;
$count \leftarrow (count + 1)$;
$j \leftarrow count$\;
\par \If {$b_7 == 0 \ and \ b_0 == 1$} {   $\rhd$ local SDN controller\

            \If {$j \leq 255$} {
                $IP_N \leftarrow b_{15}.b_{14}.b_{13}.b_{12}.b_{11}.b_{10}.b_9.b_8.j.b_6.b_5.b_4.b_3.b_2.b_1.b_0$\;
            }
            \Else {
                $count1 \leftarrow (count1 + 1)$;
                $i \leftarrow count1$\;
                \If {$i \leq 255$} {
                        $IP_N \leftarrow b_{15}.b_{14}.b_{13}.b_{12}.b_{11}.b_{10}.b_9.b_8.b_7.b_6.b_5.b_4.b_3.b_2.b_1.i$\;
                }
            }
            %\Return $(NIP)_{HEX}$\;
}
\Else { $\rhd$  Other IoT devices acting as proxies\\
            \If {$j \leq 255$} {
                \uIf {$b_7 == 0  \ and \ b_0 \neq 1$} {
                    $IP_N \leftarrow b_{15}.b_{14}.b_{13}.b_{12}.b_{11}.b_{10}.b_9.b_8.j.b_6.b_5.b_4.b_3.b_2.b_1.b_0$\;
                }
                \uElseIf {$b_7 \neq 0 \ and \ b_6 == 0$} {
                    $IP_N \leftarrow b_{15}.b_{14}.b_{13}.b_{12}.b_{11}.b_{10}.b_9.b_8.b_7.j.b_5.b_4.b_3.b_2.b_1.b_0$\;
                }
                \uElseIf {$b_6 \neq 0 \ and \ b_5 == 0$} {
                    $IP_N \leftarrow b_{15}.b_{14}.b_{13}.b_{12}.b_{11}.b_{10}.b_9.b_8.b_7.b_6.j.b_4.b_3.b_2.b_1.b_0$\;
                }
                \uElseIf {$b_5 \neq 0 \ and \ b_4 == 0$} {
                    $IP_N \leftarrow b_{15}.b_{14}.b_{13}.b_{12}.b_{11}.b_{10}.b_9.b_8.b_7.b_6.b_5.j.b_3.b_2.b_1.b_0$\;
                }
                \uElseIf {$b_4 \neq 0 \ and \ b_3 == 0$} {
                    $IP_N \leftarrow b_{15}.b_{14}.b_{13}.b_{12}.b_{11}.b_{10}.b_9.b_8.b_7.b_6.b_5.b_4.j.b_2.b_1.b_0$\;
                }
                \uElseIf {$b_3 \neq 0 \ and \ b_2 == 0$} {
                    $IP_N \leftarrow b_{15}.b_{14}.b_{13}.b_{12}.b_{11}.b_{10}.b_9.b_8.b_7.b_6.b_5.b_4.b_3.j.b_1.b_0$\;
                }
                \ElseIf {$b_2 \neq 0 \ and \ b_1 == 0$} {
                    \uIf {$b_2 == 255 \ and \ b_0 == 255 \ and \ j == 255$} {
                        $b_0 = 254$\;
                        $IP_N \leftarrow b_{15}.b_{14}.b_{13}.b_{12}.b_{11}.b_{10}.b_9.b_8.b_7.b_6.b_5.b_4.b_3.b_2.j.b_0$\;
                    }
                    \Else {
                        $IP_N \leftarrow b_{15}.b_{14}.b_{13}.b_{12}.b_{11}.b_{10}.b_9.b_8.b_7.b_6.b_5.b_4.b_3.b_2.j.b_0$\;
                    }
                }
            }

}
\Return $(IP_N)_{HEX}$\;
}\caption{ip-generation()\label{GENERATEUNIQUEIP}}%\_unique\_ip()}\label{GENERATEUNIQUEIP}}
\end{function}

We assume that the global SDN controller runs an addressing application to configure all the local SDN controllers in the many different IoT application domains. Each local SDN controller also runs the proposed addressing application to configure any SDN and IoT devices in its domain. We further assume that a local SDN controller is configured with an IP address, say CEDF:0CB8:8BA3:8A2E::0001, by the global SDN controller. In our context, CEDF:0CB8:8BA3:8A2E is the network domain and 0000:0000:0000:0001 is the identifier of the local SDN controller. The local SDN controller can assign the network prefix CEDF:0CB8:8BA3:8A2E and the device identifiers ranging from 1.0.0.0.0.0.0.1 to 255.0.0.0.0.0.0.1 and from 0.0.0.0.0.0.0.2 to 0.0.0.0.0.0.0.255 to IoT devices in the domain. 
 
In our example, the IoT device that has host identifier 0.0.0.0.0.0.0.2 and a proxy with host identifier 0.0.0.0.0.0.0.255 can allocate addresses from 0.0.0.0.0.0.1.2 to 0.0.0.0.0.0.255.2 and addresses from 0.0.0.0.0.0.1.255 to 0.0.0.0.0.0.255.255 in the dotted decimal format (\textsl{DEC}), respectively. Therefore, one can easily see that a node with host identifier 0.255.255.255.255.255.255.255 can assign addresses in the range between 1.255.255.255.255.255.255.255 and 255.255.255.255.255.255.255.254, with a network prefix of CEDF:0CB8:8BA3:8A2E.

\begin{figure}[!ht]
\centering 
\includegraphics[height=0.6\textwidth, width=0.7\textwidth]{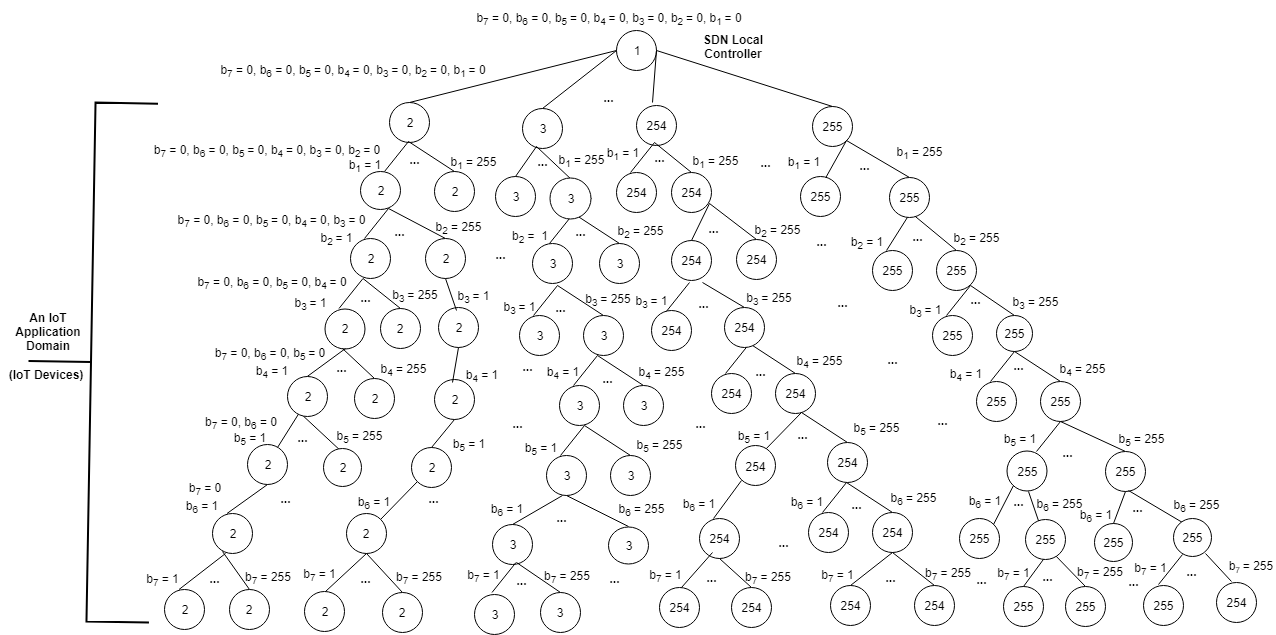}
\caption{Address allocation tree in the SDN-IoT based smart city: A simplified example} 
\label{Alloctree}
\end{figure}

Figure \ref{Alloctree} describes a simple example of how a peer or neighboring IoT device can allocate unique address (i.e. acting as a \emph{proxy}), where the last byte ($b0$) of an IP address is presented within the circle and the remaining bytes ($b_7, b_6, b_5, b_4, b_3, b_2, b_1$) outside the circle. In the event that a proxy (i.e. the IoT device) does not have available IP address for nodes that have just joined the infrastructure, then this particular proxy will need to request for new IP address(es) from their parent device. Similarly, in the unlikely event that the parent device does not have any available IP address for allocation, then a similar request will be made to the parent of this particular parent device. This allows the network to scale easily. Thus, in our proposed addressing scheme, address can be  uniquely allocated from $b_{15}.b_{14}.b_{13}.b_{12}.b_{11}.b_{10}.b_9.b_8.0.0.0.0.0.0.0.1$ to $b_{15}.b_{14}.b_{13}.b_{12}.b_{11}.b_{10}.b_9.b_8.255.255.255.255.255.255.255.254$ in the network.

We also remark that in our proposed addressing scheme, the \emph{allocation status} is maintained by the individual device. Such a status records the last assigned address (i.e. \emph{count} value), to avoid proxy devices from generating the same IP address. This allows us to avoid the need to introduce complex duplicate address detection mechanism during the process of address resolution. Further, new device obtains an IP address from its neighbor; therefore, the proposed scheme has minimal addressing overhead and latency.

\begin{table}
\center
\caption{Comparison of Address Allocation Approaches in Smart Cities}
\label{comparison1}
\begin{tabular}{@{}ccccccccccc@{}}
\hline
{Scheme} &{IP}  &{Uniqueness}  &{Addressing} &{Addressing} &{Scalability} &{Complexity}\\

{}       &  {Family} &          & {Latency}  &{Overhead}   &      & \\
\hline
{DHCP}   & IPv4, IPv6        &{Yes}    &{$\textsl{O(2*t*d)}$}  &\textsl{O($n^{2}$)}    &{Low}       &{Low}        \\
\hline
{DAD}   & IPv4, IPv6        &{No}    &{$\textsl{O(2*t*d)}$}  &\textsl{O($n^{2}$)}    &{Low}       &{Medium}        \\
\hline
{Proposed}   &IPv6        &{Yes}    &{$\textsl{O(2*t)}$}  &\textsl{O($2*l/n$)}    &{High}       &{Low}        \\
\hline
\end{tabular}
\end{table}

\subsection{Performance Evaluation} Table \ref{comparison1} compares the proposed address allocation scheme between the traditional DHCP and DAD schemes. Here, $n$ be the total number of IoT devices, $l$ the average number of links between devices, $d$ the network diameter and $t$ be the average 1-hop latency. We consider the following parameters to analyze the performance of our proposed addressing scheme along with DHCP and DAD schemes:

\noindent \textit{Uniqueness:} The most important metrics in address allocation scheme is to guarantee the uniqueness of the allocated addresses of each device. This unique address is needed to identify the device uniquely and also for unicast communication, and routing in a smart city. DAD does not guarantee the uniqueness of the allocated address whereas the proposed scheme and DHCP provide unique address allocation to each IoT device.

\noindent \textit{Addressing Latency:} This parameter is the time difference between points when a new device sends the request for an address and when it receives the address from the network. In DHCP, the new device needs to discover the DHCP server where an address request message is flooded in the whole network. The DHCP server sends the address to the new device in response. Therefore, the addressing latency of DHCP is O($2*t*d$). In DAD, the new device floods an address request message in the whole network and sets a timer based on the diameter of the network for receiving the address reply message. The new device configures itself when the timer expires. Thus, the addressing latency of DAD is O($2*t*d$). Whereas the new device acquires an address from a neighbour in our proposed addressing scheme. Therefore, the addressing latency of the proposed scheme is O($2*t$).         

\noindent \textit{Addressing Overhead:} Addressing overhead of an addressing protocol refers to the average number of messages required for an address allocation to a new device. In DHCP, the new device floods a message throughout the smart city to discover the DHCP server. Therefore, the addressing overhead of DHCP is O($n^{2}$). In DAD, the new device randomly picks a temporary address and floods a message in the whole smart city network. Therefore, the addressing overhead of DAD considered to be O($n^{2}$). In our proposed scheme, the new device obtains an address from one of its neighbours, thus the addressing overhead is O($2*l/n$).     

\noindent \textit{Scalability:} The scalability of an addressing scheme is considered to be high if the scheme does not degrade much its performance with respect to addressing latency and overhead even when the size of the network is large. The addressing overhead and the addressing latency of DHCP and DAD schemes are O($n^{2}$) and O($2*t*d$) respectively. Therefore, these schemes are considered to be low scalability. Whereas the proposed addressing scheme is considered to be highly scalable as it has O($2*l/n$) and O($2*t$) as the addressing overhead and latency respectively.

\noindent \textit{Complexity:} The addressing scheme should use the network resources (e.g., energy and memory of IoT devices, network-bandwidth) as minimal as possible at the time of address allocation. The complexity of DAD scheme is considered to be medium as it generates address from a random number and assigns to a new device. Whereas the proposed addressing scheme has low complexity as it does not need to maintain the address blocks and complex functions to generate addresses. In the proposed scheme, the existing devices (already configured with addresses) in the network acting as proxies and capable of generating addresses for new devices. This reduces the complexity and memory requirement of the proposed scheme even further.

\section{Conclusion}\label{conclu}
In this Chapter, we proposed an SDN-IoT-based smart city framework, and a distributed IPV6-based address allocation scheme. In the latter, each device in the city acts as a proxy and is capable of assigning IP addresses to new devices dynamically. We explained how the proposed approach achieves  bandwidth and energy savings in IoT devices, as well as having low addressing overhead and latency since new devices obtain their addresses from their neighbors.